\documentclass[preprint,preprintnumbers,amsmath,amssymb,
superscriptaddress,floatfix]{revtex4}
\usepackage{graphicx}
\usepackage{amsmath}
\usepackage{bm}
\usepackage{mathrsfs}
\usepackage{color}
\usepackage{slashed}
\usepackage{dcolumn}
\usepackage[normalem]{ulem}

\newcommand{\be}{\begin{equation}}
\newcommand{\ee}{\end{equation}}
\newcommand{\ba}{\begin{eqnarray}}
\newcommand{\ea}{\end{eqnarray}}

\begin{document}

\title{Does $f_1(1420)$ have a double-peak structure?}

\author{Bao-Xi Sun}
\email{sunbx@bjut.edu.cn}
\affiliation{School of Physics and Optoelectronic Engineering, Beijing University of Technology, Beijing 100124, China}

\date{\today}

\begin{abstract}
The one-pion exchange interaction between the kaon and the vector antikaon is investigated by solving the Schrodinger equation in the S-wave approximation. In addition to the $f_1(1285)$ particle, another
bound state of $K \bar{K}^*$ is found, which is approximately 9 MeV below the threshold of $K \bar{K}^*$ and labeled $f_1(1378)$ for convenience in this manuscript. Under the outgoing wave condition, two resonance states of $K \bar{K}^*$  are produced by solving the Schrodinger equation with different coupling constants in the one-pion-exchange potential fixed with the binding energies of $K \bar{K}^*$ bound states $f_1(1285)$ and $f_1(1378)$ respectively. Both of the resonance states are located in the vicinity of 1400 MeV, and thus it is reasonable to assume that these two resonance states correspond to the $f_1(1420)$ particle in the review of the Particle Data Group although they arise from different couplings of the kaon and vector antikaon, respectively.
\end{abstract}

\maketitle

\section{Introduction}
\label{sect:Introduction}

The $K \bar{K}^*$ system is investigated by solving the Schrodinger equation under the outgoing wave condition, and a complex solution above the threshold of $K \bar{K}^*$ is obtained, which is assumed to be the $f_1(1420)$ particle in the review of the Particle Data Group(PDG)\cite{PDG2024}. Reasonably, this method is extended to study the $D \bar{D}^*$ case\cite{sunbx2024}. When the Schrodinger equation is solved, an one-pion-exchange potential of the kaon and vector antikaon is adopted, which is different from the kernel used in the unitary coupled channel approximation, where a vector meson exchange is dominant according to the SU(3) hidden guage 
symmetry\cite{Roca2005,sun1420}.

The vector meson is included as a gauge boson in the Lagrangian density when the hidden gauge symmetry in the SU(3) flavor space is taken into account.  
Therefore, the interaction of vector mesons can be obtained, and the bound and resonance states related to the vector meson- vector meson interaction are generated dynamically when the Bethe-Salpeter equation is solved in the unitary coupled-channel approximation\cite{Molina,Geng}. Although vector meson exchange plays a dominant role in their calculation, these resonance states produce a width when the mass distribution of the vector meson is considered, especially when the $\pi \pi $ decay channel is taken into account via a box diagram. Apparently, pion exchange is essential in the generation of resonance states. However, when the vector meson and baryon interaction is studied, the pion exchange via the box diagram is excluded in the calculation since a more intelligent scheme is proposed. In similarity with the pseudoscalar meson and baryon interaction, the $t-$ channel interaction of vector mesons and baryons can be obtained conveniently via a vector meson exchange between the vector meson and the baryon\cite{Ganzalez,Sarkar,Ramos}. Along this clue, the $K\bar{K}^*$ interaction can also be evaluated via a vector meson exchange, and the one-pion-exchange potential between the kaon and the vector antikaon is neglected when the Bethe-Salpeter equation is solved\cite{sun1420}. 
Actually, the pion mass is far less than the mass of vector mesons, and the range of force is larger than the radius of the nucleon when a pion transfers between the vector meson and the baryon. However, the range of force is less than the radius of the nucleon even if the lightest vector meson exchange is considered.

When the interaction between the vector meson and the baryon is evaluated in the vicinity of the threshold, the momentum of the intermediate vector meson can be neglected, and the propagator becomes a reciprocal of the mass square of the vector meson, which implies that the $t-$ channel interaction reduces to a contact term. In this case, it can play the role of a kernel when the Bethe-Salpeter equation is solved in the unitary coupled-channel approximation.
In other words, the intermediate meson propagator in the $t-$ channel interaction must be replaced with a constant, and then it can be used as a kernel of the Bethe-Salpeter equation. In the unitary coupled-channel approximation, only a potential between the vector meson and the baryon is needed. Whether or not the kind of intermediate meson is changed is not important.

In this work, the one-pion-exchange interaction between the kaon and the vector antikaon is taken into account, and the $K\bar{K}^*$ system is investigated by solving the Schrodinger equation in the S-wave approximation. Consequently, the possible $K\bar{K}^*$ bound and resonance states correspond to solutions of the Schrodinger equation under different boundary conditions of the wave function, respectively.  
In addition to the $f_1(1285)$ particle, which is regarded as a bound state of $K \bar{K}^*$ with a binding energy of 105 MeV, another bound state of $K \bar{K}^*$ approximately 9 MeV below the threshold of $K \bar{K}^*$ is predicted and labeled $f_1(1378)$ for convenience in this manuscript. Moreover, two resonance states of $K \bar{K}^*$ are found by solving the Schrodinger equation under the outgoing wave condition, which are related to two
bound states of $K \bar{K}^*$, $f_1(1285)$ and $f_1(1378)$ particles, respectively.

In the next section, descriptions of the Schrodinger equation with an one-pion-exchange potential will be reviewed according to Ref.~\cite{sunbx2024}. In Section~\ref{sect:KKstar}, the calculation results for the bound and resonance states of $K \bar{K}^*$ will be discussed. Finally, Section~\ref{sect:summary} is devoted to a brief summary. 

\section{Framework}
\label{sect:framework}

If the interaction of the kaon and the vector antikaon is realized by exchanging a pion, the potential of them can be indicated as a Yukawa type, 
\be
\label{eq:202307071816}
V(r)=-g^2\frac{e^{-mr}}{d},
\ee
where $m$ is the mass of the pion, $g$ is the coupling constant, and the distance $r$ in the denominator has been replaced approximately with the force range $d=1/m$.
It is apparent that the potential in Eq.~(\ref{eq:202307071816}) is reasonable in the force range, and it is equal to the original Yukawa type potential asymptotically at infinity. Under this approximation, the Schrodinger equation can be solved analytically.

Supposing the radial wave function $R(r)=\frac{u(r)}{r}$, the radial Schrodinger equation in the S-wave approximation can be written as
\be
\label{eq:202307081218}
-\frac{\hbar^2}{2\mu} \frac{d^2 u(r)}{dr^2}+V(r)u(r)=Eu(r),
\ee
where $\mu$ is the reduced mass of the two-body system.

According to the substitution of variables 
\be 
u (r) = J (x),
\ee
and
\be
r \rightarrow x=\alpha e^{-\beta r},~~~~0 \le x \le \alpha,
\ee
with
\be
\alpha=2g \sqrt{2 \mu d},~~~~\beta=\frac{1}{2d},
\ee
and
\be
\label{eq:rhoenergy}
\rho^2=-8d^2 \mu E,~~~~E \le 0,
\ee
the radial Schrodinger equation in Eq.~(\ref{eq:202307081218}) becomes the $\rho$th order Bessel equation, 
\be
\label{eq:202307081903}
\frac{d^2 J(x)}{d x^2}+\frac{1}{x} \frac{d J(x)}{d x} +\left[1-\frac{\rho^2}{x^2}\right] J(x)=0,
\ee
and its solution is the $\rho$th order Bessel function $J_\rho(x)$.

For the bound state, when $r \rightarrow +\infty$, the radial wave function $R(r) \rightarrow 0$, which implies that $u(r)=J_\rho(\alpha e^{-\beta r})=J_\rho(0)$ with $\rho \ge 0$. On the other hand, when $r \rightarrow 0$, $u(r) \rightarrow 0$, and it means
\be
\label{eq:Besselzeropoint}
J_\rho(\alpha)=0.
\ee
Therefore, if only one bound state of $K\bar{K}^*$ has been detected, the order of the Bessel function $\rho$ in Eq.~(\ref{eq:Besselzeropoint}) can be determined with the binding energy $E$, and then the coupling constant $g$ in the Yukawa potential is obtained with the first nonzero zero point of the Bessel function $J_\rho(\alpha)$, which takes a form of
\be
\label{eq:couplingzeropoint}
g^2=\frac{\alpha^2}{8 \mu  d}.
\ee

The first zero point of the zeroth order Bessel function $J_0(\alpha)$ lies at $\alpha=2.405$, while the second and third zero points of it lie at $\alpha=5.520$ and $\alpha=8.654$, respectively. 
The solution of Eq.~(\ref{eq:Besselzeropoint}) corresponds to the non-zero zero point of the $\rho$th-order Bessel function $J_\rho(\alpha)$. 
If the nonzero zero point of the $\rho$th order Bessel function $J_\rho(\alpha)$ is greater than the first zero point of the zeroth order Bessel function $J_0(\alpha)$ and less than the second zero point of $J_0(\alpha)$, that is, it belongs to the range of $\left[2.405,5.520\right]$, there exists only one bound state of $K\bar{K}^*$. However, if the solution of Eq.~(\ref{eq:Besselzeropoint}) lies in the range of $[5.520,8.654]$, it is the zero point of two different order Bessel functions, which implies that there might be two bound states of $K\bar{K}^*$.

The Hankel functions $H_\rho^{(1)}(x)$ and $H_\rho^{(2)}(x)$ are also two independent solutions of the Bessel equation. Under the outgoing wave condition
\be
\label{eq:outgoing}
H_\rho^{(2)}(\alpha)=0,
\ee
the $K\bar{K}^*$ resonance state can be obtained when the value
$\alpha$ is fixed with Eq.~(\ref{eq:Besselzeropoint}).
Apparently, the energy of the resonance state is related to the order of the second kind of Hankel function and takes a complex value
of $E=M-i\frac{\Gamma}{2}$, where the real part represents the mass of the resonance state, while the imaginary part is one half of the decay width, i.e. $\Gamma=-2iImE$, as discussed in Ref.~\cite{Moiseyev}.

\section{Results}
\label{sect:KKstar}

\begin{figure}
\includegraphics[width=0.7\textwidth]{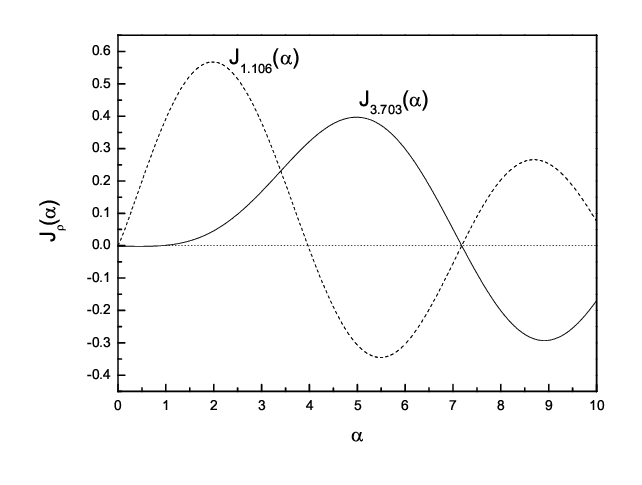}
\caption{
Bessel functions $J_\rho(\alpha)$ with $\rho=3.703$(Solid line) and $\rho=1.106$(Dash line) are depicted respectively.
}
\label{fig:Bessel}
\end{figure}

The $K \bar{K}^*$ interaction has been studied by solving the Schrodinger equation in the S-wave approximation, and a $K \bar{K}^*$ bound state and a $K \bar{K}^*$ resonance state are obtained as solutions of the Schrodinger equation under different boundary conditions of the wave function, which are assumed to be counterparts of the $f_1(1285)$ and $f_1(1420)$ particles listed in the PDG data, respectively\cite{sunbx2024}.

The particle $f_1(1285)$ is treated as a deep bound state of $K\bar{K}^*$ with a binding energy of $105$MeV, so the Bessel function, which appears as a solution of the Schrodinger equation in the Yukawa type potential, has a zero point of $\alpha=7.1831$.
Its value is larger than the second nonzero zero point of the zeroth order Bessel function $J_0(\alpha)$ so that there exists another $K\bar{K}^*$ bound state besides the $f_1(1285)$ particle.
Since the Bessel function $J_\rho(\alpha)$ with an order of $\rho=1.106$ is zero in $\alpha=7.1831$, as shown in Fig.~\ref{fig:Bessel}, the binding energy of another $K\bar{K}^*$ bound state can be determined according to Eq.~(\ref{eq:rhoenergy}), which is about $9$MeV lower than the threshold of $K\bar{K}^*$, and has the same spin and parity as those of the particle $f_1(1285)$, which is labeled $f_1(1378)$ for convenience in this manuscript.
This weak bound state near the $K\bar{K}^*$ threshold has not been listed in the PDG manual, and thus it would be the task of experimental scientists to discover it in the future.

\begin{figure}
\includegraphics[width=0.45\textwidth]{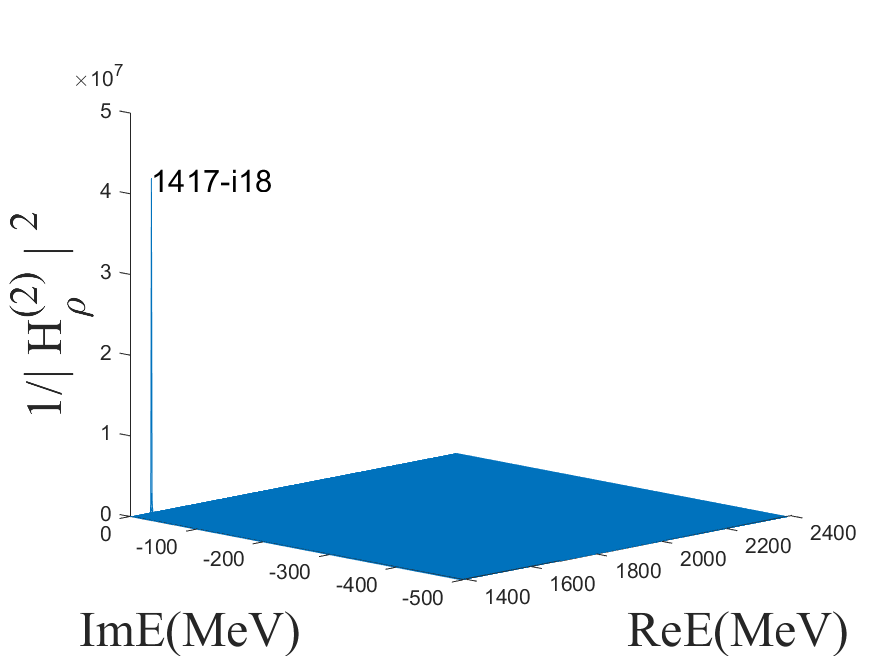}
\includegraphics[width=0.45\textwidth]{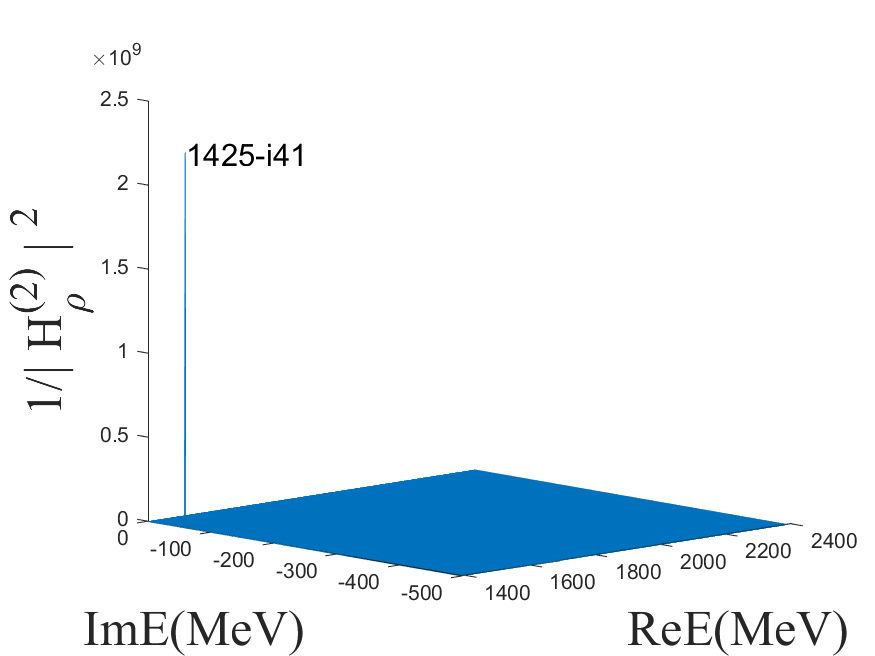}
\caption{$1/|H_\rho^{(2)}(\alpha)|^2$ .vs. the complex energy $E$ with $\alpha=7.1831$(left) and $\alpha=3.9696$(right)  in the $K \bar{K}^*$ case. The pole of $1/|H_\rho^{(2)}(\alpha)|^2$ corresponds to a zero-point of the second kind of Hankel function $H_\rho^{(2)}(\alpha)$, which represents a resonance state of $K \bar{K}^*$, as labeled in the figure.}
\label{fig:kkstar}
\end{figure}

The first nonzero zero point of the Bessel function $J_{1.106}(\alpha)$ lies at $\alpha=3.9696$, while $\alpha=7.1831$ is the second nonzero zero point of it. Undoubtedly, the resonance state of $K\bar{K}^*$ can also be determined according to the outgoing wave condition in Eq.~(\ref{eq:outgoing})
with $\alpha=3.9696$. Along this clue, another $K\bar{K}^*$ resonance state is found, which lies at $1425-i41$MeV on the complex energy plane in the center-of-mass frame. Apparently, its mass and width are close to those of the resonance state obtained with the zero point $\alpha=7.1831$, respectively, as shown in Fig.~\ref{fig:kkstar}, so it can be assumed that these two resonance states obtained with different zero points correspond to the same particle $f_1(1420)$ in the PDG data. The energies of the $K\bar{K}^*$ resonance states and the properties of the $f_1(1420)$ particle are summarized in Table~\ref{table:KKstar}.

According to Eq.~(\ref{eq:couplingzeropoint}), different zero point values correspond to different coupling constant $g$ in the Yukawa type of potential. The coupling constant $g=1.682$ is obtained with the zero point at $\alpha=7.1831$, while a more weaker coupling constant $g=0.9293$ is related to the first non-zero zero point at $\alpha=3.9696$ of the Bessel function $J_{1.106}(\alpha)$. 
\begin{table}[htbp]
 \renewcommand{\arraystretch}{1.2}
\centering
\vspace{0.5cm}
\begin{tabular}{c|c|c|c|c|c}
\hline\hline
 $\alpha$ & energy(MeV) &   name & $I^G(J^{PC})$ &  mass(MeV) & width(MeV) \\
 \hline
$7.1831$ & $1417-i18$ & $f_1(1420)$ & $0^+(1^{++})$ &   $1428.4^{+1.5}_{-1.3}$  & $56.7\pm3.3$
  \\
$3.9696$ & $1425-i41$ &  &  &     & 
  \\  
\hline\hline
\end{tabular}
\caption{
The complex energy of the possible $K \bar{K}^*$ ($\bar{K} K^*$) resonance state with the outgoing wave condition in Eq.~(\ref{eq:outgoing}) and the possible counterpart in the PDG data.
}\label{table:KKstar}
\end{table}

\section{Summary}
\label{sect:summary}

The one-pion-exchange interaction between the kaon and the vector antikaon is investigated by solving the Schrodinger equation in the S-wave approximation. If the wave function vanishes at the infinite, the solution of the Schrodinger equation takes the form of the Bessel function, which corresponds to a bound state of the $K \bar{K}^*$ system. Since the $f_1(1285)$ particle is 105 MeV lower than the threshold of $K \bar{K}^*$, the coupling constant in the one-pion-exchange interaction can be determined according to the first nonzero zero point of the corresponding Bessel function. Therefore, a resonance state of $K \bar{K}^*$ can be obtained as a solution of the Schrodinger equation under the outgoing condition of the wave function, which lies at $1417-i18$MeV on the complex energy plane in the center of mass frame.  
However, the first nonzero zero point of the Bessel function $J_{3.703(\alpha)}$ appears at $\alpha=7.1831$ when the $f_1(1285)$ particle is treated as a bound state of $K \bar{K}^*$, which is above the second zero point of the zeroth order Bessel function at $\alpha=5.520$. It is apparent that the value $\alpha=7.1831$ is the zero point of another Bessel function in addition to the Bessel function $J_{3.703(\alpha)}$, and thus there exists another bound state of $K \bar{K}^*$ about 9 MeV below the threshold of $K \bar{K}^*$, which is not listed in the PDG data. 
These two bound states of $K \bar{K}^*$ are obtained as solutions of the Schrodinger equation in the S-wave approximation, and the coupling constant in the one-pion exchange potential of the $K \bar{K}^*$ system can be determined with their bound energies, respectively, whose values are different from each other. It implies that the interaction of between the kaon and the vector antikaon might be more complicated than a simply Yukawa type potential. 
Fortunately, these resonance states of $K \bar{K}^*$ both appear around 1400 MeV on the complex energy plane, and it is reasonable to assume that both correspond to the same $f_1(1420)$ particle listed in the PDG data. However,  since the coupling constant in the $K \bar{K}^*$ interaction is different from each other, these two resonance states are produced differently. In other words, the $f_1(1420)$ particle has a double-peak structure caused by the strange character of the $K \bar{K}^*$ interaction. 
%



\begin{thebibliography}{99}


\bibitem{PDG2024}
S.~Navas \textit{et al.} [Particle Data Group],
``Review of particle physics,''
Phys. Rev. D \textbf{110}, no.3, 030001 (2024)



\bibitem{sunbx2024}
B.~X.~Sun, Q.~Q.~Cao and Y.~T.~Sun,
``The possible KK\textasciimacron{}* and DD\textasciimacron{}* bound and resonance states by solving the Schrodinger equation,''
Commun. Theor. Phys. \textbf{76}, no.10, 105301 (2024)

\bibitem{Roca2005}
L.~Roca, E.~Oset and J.~Singh,
``Low lying axial-vector mesons as dynamically generated resonances,''
Phys.\ Rev.\ D {\bf 72}, 014002 (2005)

\bibitem{sun1420}
D.~M.~Wan, S.~Y.~Zhao and B.~X.~Sun,
``The $K\bar{K}^*$ interaction in the unitary coupled-channel approximation,''
arXiv:1808.08358 [hep-ph].

\bibitem{Molina}
R.~Molina, D.~Nicmorus and E.~Oset,
``The rho rho interaction in the hidden gauge formalism and the f(0)(1370) and f(2)(1270) resonances,''
Phys. Rev. D \textbf{78}, 114018 (2008)

\bibitem{Geng}
L.~S.~Geng and E.~Oset,
``Vector meson-vector meson interaction in a hidden gauge unitary approach,''
Phys. Rev. D \textbf{79}, 074009 (2009)

\bibitem{Ganzalez}
P.~Gonzalez, E.~Oset and J.~Vijande,
``An Explanation of the Delta(5/2-)(1930) as a rho Delta bound state,''
Phys. Rev. C \textbf{79}, 025209 (2009)

\bibitem{Sarkar}
S.~Sarkar, B.~X.~Sun, E.~Oset and M.~J.~Vicente Vacas,
``Dynamically generated resonances from the vector octet-baryon decuplet interaction,''
Eur. Phys. J. A \textbf{44}, 431-443 (2010)

\bibitem{Ramos}
E.~Oset and A.~Ramos,
``Dynamically generated resonances from the vector octet-baryon octet interaction,''
Eur. Phys. J. A \textbf{44}, 445-454 (2010)

\bibitem{Moiseyev}
N. Moiseyev, {\it Non-Hermitian Quantum Mechanics}, Cambridge University Press, New York, 2011.

\end{thebibliography}
\end{document}